\documentclass[11pt]{article}

\usepackage[utf8]{inputenc}
\usepackage[T1]{fontenc}

\usepackage{amsmath,amssymb}
\usepackage{amsthm}

\theoremstyle{plain}
\newtheorem{lemma}{Lemma}
\newtheorem{theorem}{Theorem}

\theoremstyle{remark}
\usepackage{tikz}
\usetikzlibrary{positioning}
\usetikzlibrary{arrows.meta}

\usepackage{graphicx}
\usepackage{caption}
\usepackage{pgfplots}
\pgfplotsset{compat=1.18}
\usepgfplotslibrary{groupplots}
\usepackage[table]{xcolor}
\usepackage{array}
\usepackage{geometry}
\usepackage{makecell}
\usepackage{adjustbox}
\usepackage{pdflscape}
\usepackage{booktabs}
\usepackage{amsthm}

\usepackage{float}
\usepackage[verbose]{hyperref}

\begin{document}

\markboth{M. Nowak-Kępczyk}{Frobenius Revivals...}
\title{Frobenius Revivals in Laplacian Cellular Automata: Replication, Entropy, and Information Recovery}
\author{Małgorzata Nowak-Kępczyk	\footnote{Institute of Informatics, John Paul II Catholic University of Lublin, 
		20-708 Lublin, Poland,\\
		malnow@kul.pl}}

\date{}

\maketitle{}

\begin{abstract}
We investigate Frobenius revivals in prime-modulus
Laplacian cellular automata. The Frobenius identity
yields exact superpositions of translated,
coefficient-weighted seed copies at prime-power times;
we distinguish this algebraic replication from their
geometric separation.

We investigate seed recovery from separated replicas
and evaluate the correction of independent symbol errors
introduced during transmission of encoded configurations.

For the diagonal neighborhood, we determine the
exact periods of moving corner operators on square
windows and establish sufficient conditions for recovering
any binary seed under a binary--ternary--binary schedule.
The results connect prime-power algebraic structure with
spatial pattern formation and precisely delimited forms
of information recovery.

\end{abstract}




\vspace{1em}

\section*{Highlights}
\begin{itemize}
	\item Frobenius identities yield exact seed replication
	at prime-power times in Laplacian cellular automata.
	
	\item Spatial separation of replicas depends on
	seed geometry, neighbourhood mask, and modulus.
	
	\item Entropy dynamics reveal pronounced changes
	in spatial organisation near replication events.
	
	\item Separated replicas enable exact noiseless
	information recovery and support consensus-based
	reconstruction under noise.
	
	\item Exact corner periods yield guaranteed seed recovery
	under selected binary--ternary--binary schedules.
\end{itemize}

\section{Introduction}

Cellular automata provide a simple framework for studying
the emergence of complex spatial and temporal structures
from local update rules. Even linear cellular automata
may generate configurations of considerable visual
complexity, with irregular patterns interrupted by
strikingly ordered arrangements. Understanding how
such organisation arises is a natural question at the
intersection of discrete dynamics, algebra, and
computation.

A particularly transparent mechanism occurs in modular
discrete Laplacian dynamics. When the modulus is prime,
the Frobenius identity simplifies high powers of the
evolution operator into finite sums of spatial
translations. Consequently, a configuration that has
undergone a long dispersive evolution may reorganise
at a prime-power time into exact, coefficient-weighted
copies of its initial seed. Although the intermediate
patterns may appear irregular, the replication event
is determined entirely by algebraic properties of
the update rule.

This phenomenon raises several related questions.
How does the geometry of the initial seed influence
the appearance of separated replicas? How are
replication events reflected in the entropy of the
evolving configurations? Can the resulting spatial
redundancy be used to recover information from
intermediate or perturbed states?

In this work, we examine these questions for
two-dimensional Laplacian cellular automata.
We establish the algebraic replication formula
and distinguish exact replication from the
geometric separation of individual copies.
Numerical experiments illustrate the evolution
of spatial patterns and Shannon entropy under
different prime moduli.

We further investigate information recovery
based on Frobenius replication. In the noiseless
constant-prime setting, an initial configuration
can be reconstructed by continuing the evolution
to a suitable replication time and extracting
a complete, appropriately normalised replica.
We also examine reconstruction under local
perturbations and accumulated noise, using
replica consensus as a computational procedure.

Finally, we consider evolutions in which the
modulus changes over time. Such mixed-modulus
dynamics are sensitive to the order and placement
of individual modular stages, offering an
additional means of modifying the resulting
spatial structures. Unlike fixed-prime evolution,
however, their replication properties do not
follow directly from a single Frobenius identity.

Our aim is therefore to connect an exact
algebraic replication mechanism with its
observable dynamical signatures and potential
use in deterministic information recovery,
while distinguishing established properties
from computational observations and possible
extensions.

\section{Background and Previous Work}
\subsection{Related work}
The study of complex behaviour in discrete dynamical systems spans nonlinear dynamics, symbolic dynamics, and the extensive theory of cellular automata (CAs). Foundational work in chaos theory \cite{ott2002chaos,strogatz2018nonlinear} has established a framework for describing transitions between ordered and disordered regimes, including bifurcation cascades, intermittent chaotic windows, and rapid returns to regular behaviour. These concepts have been adapted to spatially extended systems, where spatiotemporal chaos and emergent structures play a central role \cite{politi1992periodic,crutchfield1989inferring}.

Within CA research, Wolfram’s pioneering programme \cite{wolfram1983statistical} revealed that simple local update rules can generate dynamics characteristic of nonlinear systems, such as long chaotic transients, spatial disorder, and spontaneous formation of coherent structures. Subsequent developments in symbolic dynamics have provided a formal perspective on CA evolution through shift spaces, subshifts of finite type, and algebraic constraints \cite{fuks2004symbolic}, clarifying how periodic orbits and low-entropy motifs arise within globally complex behaviour.

Related phenomena have also been analysed in coupled map lattices \cite{kaneko1993theory}, where coherent domains, intermittent patterns, and multi-scale organisation emerge from interactions between local rules and global phase–space structure. These systems highlight how discrete update mechanisms can generate pattern formation comparable to that observed in continuous media, paralleling behaviours seen in Laplacian CAs.

Recent work on linear and algebraic CAs has examined periodicity, invertibility, and spectral properties as determinants of long-term dynamics. Yet the route by which dispersive, seemingly chaotic evolution reorganises into large-scale ordered structure remains poorly understood, particularly in modular Laplacian settings. Although the algebraic identities governing such operators are known, their dynamical consequences—most notably multi-tile replication at prime–power times—have received limited systematic analysis.

The present study situates Frobenius-driven revivals within this broader context. Rather than emphasising pseudorandomness, diffusion, or local complexity, we focus on the global algebraic mechanism that enforces transitions from high-entropy states to sharply ordered replica configurations. This viewpoint connects algebraic CA theory with dynamical phenomena usually associated with nonlinear systems, offering a unified interpretation of chaos–order transitions in discrete computational media.

We use the term “chaotic” in the operational sense of high-entropy, visually disordered, and mixing dynamics, without implying sensitive dependence in the classical Lyapunov sense.

\subsection{Mathematical Background}

Let $p$ be prime and let $M\subset\mathbb Z^2\setminus\{0\}$
be a finite neighbourhood mask of cardinality $d$.
We consider finitely supported configurations
$u:\mathbb Z^2\to\mathbb F_p$ evolving under the
discrete Laplacian
\[
(L_Mu)(x)=\sum_{a\in M}u(x-a)-d\,u(x).
\]
The Moore and von Neumann masks are standard examples.
Writing $S_a u(x)=u(x-a)$, we have
$L_M=\sum_{a\in M}S_a-dI$.

The characteristic feature of prime-modulus evolution
follows from the Frobenius identity. Since translations
commute, for every integer $m\geq1$,
\[
L_M^{p^m}
=\sum_{a\in M}S_{p^m a}-dI
\qquad\text{over }\mathbb F_p.
\]
Thus a large iterate of the Laplacian reduces to a
finite sum of widely separated translations, with
coefficients determined by the original mask.
This explains the abrupt reappearance of ordered
configurations after long dispersive transients.

\begin{lemma}[Frobenius replication]
	\label{lem:frobenius}
	Let $u_0$ have finite support $A$, and let
	$u_t=L_M^t u_0$ over $\mathbb F_p$.
	For every $m\geq1$,
	\[
	u_{p^m}
	=\sum_{a\in M}S_{p^m a}u_0-d\,u_0.
	\]
	If the translated supports corresponding to the
	nonzero coefficients on the right-hand side are
	pairwise disjoint, then $u_{p^m}$ consists of
	spatially separated copies of the initial
	configuration, multiplied by their respective
	coefficients in $\mathbb F_p$.
\end{lemma}

\begin{proof}
	The operator $L_M$ is a finite sum of commuting
	translations and a scalar multiple of the identity.
	Applying the Frobenius identity in characteristic
	$p$ gives
	\[
	\left(\sum_{a\in M}S_a-dI\right)^{p^m}
	=\sum_{a\in M}S_a^{p^m}-d^{p^m}I.
	\]
	Since $S_a^{p^m}=S_{p^m a}$ and
	$d^{p^m}=d$ in $\mathbb F_p$, the first identity
	follows. If the corresponding translated supports
	do not overlap, their contributions cannot cancel
	or interfere, yielding separated replicas.
\end{proof}

The separation condition depends on the size of the
seed and the distances between the translation
vectors. In particular, a Frobenius identity
guarantees the replication formula, but does not
guarantee that the copies are already disjoint
at the first prime-power time.

\begin{lemma}[Finite propagation of perturbations]
	\label{lem:lightcone}
	Let $D=M\cup\{0\}$ and let $u_t=L_M^t u_0$.
	Suppose that at iteration $s$ a perturbation $\eta$
	supported in a finite set $S$ is introduced:
	$v_s=u_s+\eta$, followed by $v_{t+1}=L_Mv_t$.
	Then for every $t\geq s$,
	\[
	\operatorname{supp}(v_t-u_t)
	\subseteq S\oplus (t-s)D,
	\]
	where $(t-s)D$ denotes the Minkowski sum of
	$t-s$ copies of $D$.
\end{lemma}

\begin{proof}
	Set $\delta_t=v_t-u_t$. By linearity,
	$\delta_{t+1}=L_M\delta_t$.
	Each update can enlarge the support only by
	translations belonging to $D$, hence
	$\operatorname{supp}(\delta_{t+1})
	\subseteq\operatorname{supp}(\delta_t)\oplus D$.
	Iteration from $\delta_s=\eta$ gives the claim.
\end{proof}

The second lemma expresses a finite propagation
speed determined by the neighbourhood mask.
It provides a geometric bound on the influence
of local perturbations, but does not by itself
imply statistical independence of errors in
different replicas.

\subsection{Replication Times and Spatial Separation}
\label{subsec:replication-times}

The Frobenius identity in Lemma~\ref{lem:frobenius}
identifies a distinguished sequence of times
$t=p^m$, $m\geq1$, which we call
\emph{Frobenius replication times}.
At these times, the evolution of any finitely
supported configuration $u_0$ satisfies
\[
u_{p^m}
=\sum_{a\in M}S_{p^m a}u_0-d\,u_0
\qquad\text{over }\mathbb F_p.
\]
After combining coincident shifts and removing
terms with zero coefficients modulo $p$, this
expression is a finite superposition of
translated, coefficient-weighted copies of $u_0$.

We distinguish two geometric situations.

\begin{itemize}
	\item \emph{Overlapping replication.}
	The translated supports are not pairwise disjoint.
	Individual contributions may interfere or cancel,
	and separate copies need not be visually recognisable.
	
	\item \emph{Separated replication.}
	The translated supports corresponding to nonzero
	coefficients are pairwise disjoint. The resulting
	configuration then consists of exact, spatially
	separated copies of $u_0$, multiplied by their
	respective coefficients in $\mathbb F_p$.
\end{itemize}

For a fixed finite seed, sufficiently large
Frobenius powers guarantee separation whenever
the surviving translation vectors are distinct.
More precisely, if
$A=\operatorname{supp}(u_0)$ and $V$ is the set
of translation vectors with nonzero coefficients,
separation holds when
\[
(A+p^m v)\cap(A+p^m w)=\varnothing
\qquad\text{for all distinct }v,w\in V.
\]
The smallest Frobenius time satisfying this
condition is called the
\emph{separated replication time} of the seed,
provided that at least two nonzero translated
contributions are present.

The same identity applies to an intermediate
configuration $u_{t_0}$:
\[
u_{t_0+p^m}=L_M^{p^m}u_{t_0}.
\]
Thus \emph{shifted replication} refers to
the appearance of translated copies of an
intermediate state rather than of the initial seed.
It is not a separate algebraic mechanism.

These replication times should not be confused
with ordinary periods of the orbit.
The Frobenius identity establishes a relation
between configurations at different times,
but does not imply that the system returns to
its original state.

\subsection{Order-Dependent Mixed-Modulus Dynamics}
\label{subsec:mixed_moduli}

We now consider evolutions in which the modulus
changes during the trajectory. Unlike the
constant-prime case, such dynamics cannot in general
be described by powers of a single linear operator
over a fixed finite field.

Let $\mathcal C$ denote the set of finitely supported
configurations $u:\mathbb Z^2\to\mathbb Z$.
For every integer $k\geq2$, define
\[
F_k(u)=R_k(L_Mu),
\]
where $L_M$ is the integer-valued discrete
Laplacian associated with a fixed neighbourhood
mask $M$, and $R_k$ denotes pointwise reduction
to the representatives $\{0,\ldots,k-1\}$.

For a prescribed modulus sequence
$\mathbf{k}=(k_1,\ldots,k_T)$, the evolution is
given by
\[
u_T=F_{k_T}\circ\cdots\circ F_{k_1}(u_0).
\]
Thus the dynamics depend not only on the moduli
used, but also on their order.

\paragraph{Order dependence.}
Although the underlying integer Laplacian is
unchanged, reductions modulo different integers
generally do not commute. In particular,
\[
F_p\circ F_q\neq F_q\circ F_p
\]
may hold for distinct primes $p$ and $q$.
Consequently, two sequences containing the same
moduli, with the same multiplicities, may produce
different configurations.

This distinguishes mixed-modulus evolution from
the constant-prime setting, where successive
updates are powers of a single linear operator.
The order of the modular stages is therefore an
essential dynamical parameter.

\paragraph{Placement relative to revival times.}
A second aspect is the position of a modular
intervention along the trajectory. Suppose that
a binary evolution is interrupted by a block of
$b$ ternary updates and subsequently returns to
modulus $2$. Such a composition may be written as
\[
F_2^{\,c}\circ F_3^{\,b}\circ F_2^{\,a},
\]
where $a$, $b$, and $c$ specify the durations of
the successive stages.

Changing these durations may alter the resulting
configuration even when their sum remains fixed.
In particular, interventions placed near the
Frobenius times of the binary dynamics can interact
with configurations that are undergoing algebraically
determined replication.

The resulting evolution need not preserve the
replication schedule of either constant-modulus
system. Depending on the intervention and its
placement, previously separated structures may
overlap, new patterns may appear, or subsequent
replication events may change their geometry.

\paragraph{Scope of the construction.}
The Frobenius identity applies exactly to blocks
of evolution over a fixed prime field. It does not,
by itself, imply an analogous identity for an
arbitrary composition of modular reductions.

Similarly, the existence of replica configurations
at selected times does not establish invertibility
of the mixed-modulus evolution or periodicity of
the complete trajectory. Such properties require
separate arguments or direct computational
verification.

Nevertheless, ordered modulus sequences provide
a flexible framework for controlling the temporal
and spatial organisation of Laplacian cellular
automata. Their study connects the algebraic
replication mechanism of constant-prime dynamics
with more general, time-dependent modular
evolution.

The influence of modular ordering and intervention
timing has also been investigated in time-dependent
Laplacian dynamics, including periodic modulus
sequences and sparse ternary interventions
\cite{MNK2026long,MNK2026clock}.

\section{Methods}
\label{sec:methods}

This section summarises the computational model, seed geometry, observables, 
noise mechanisms, and reconstruction rules used throughout the numerical study.  
The algebraic framework for Frobenius-driven revivals and the associated period 
structure was introduced in Section~2; here we focus on the experimental and 
numerical procedures.

	\subsection{Model}
	
	Numerical experiments are performed on the square
	lattice $\mathbb Z^2$, using the discrete Laplacian
	$L_M$ defined in Section~2.2. Unless stated
	otherwise, we employ the Moore neighbourhood,
	containing the eight nearest surrounding sites.
	
	For a fixed prime modulus $p$, the evolution is
	given by $u_{t+1}=R_p(L_Mu_t)$.
	For mixed-modulus experiments, the modulus may
	change at each iteration according to a prescribed
	sequence, as described in Section~2.4.
	
	All simulations use an expanding computational
	window, enlarged by one lattice site in each
	direction per iteration. This accommodates the
	finite propagation of the dynamics and avoids
	artificial interactions caused by periodic
	boundary conditions.
	
	Replication is identified by comparing the
	resulting configurations with translated copies
	of the initial seed. We distinguish algebraic
	replication from spatial separation, following
	the definitions in Section~2.3.

A \emph{seed} is a compact, nonzero pattern supported in a finite $N \times N$ window.
We use three representative classes of seeds:
\begin{itemize}
	\item small seeds (up to $3\times 3$ support),
	\item medium seeds (natural-image silhouettes, $\sim\!18\times 18$),
	\item large seeds (high-resolution shapes, $\sim\!80\times 80$).
\end{itemize}
All seeds are placed at the centre of the expanding canvas.  
Silhouettes are particularly convenient, as geometric distortions and 
misalignments among replicas are easily detected.

\subsection{Observables}

We use symbol entropy to describe the spatial organisation
of evolving states and normalised Hamming distance to
assess reconstruction accuracy.

\paragraph{Symbol entropy.}
For a configuration $u_t$ modulo $p$, let $r_t$ be the
minimal rectangular bounding box of its nonzero support.
The relative frequency of symbol $c$ within this box is
$\rho_t^{(c)}=|\{x\in r_t:u_t(x)=c\}|/|r_t|$.
We compute Shannon entropy as
$H_t=-\sum_{c=0}^{p-1}\rho_t^{(c)}\log\rho_t^{(c)}$,
with $0\log0=0$. Entropy is set to zero for an empty support.
Because the maximum is $\log p$, entropy values for
different moduli are not directly comparable without
accounting for alphabet size.

\paragraph{Reconstruction error.}
For an original seed $u_0$ and its reconstruction
$\hat u_0$ on the same $D\times D$ grid, the normalised
Hamming error is
$E=D^{-2}\sum_x\mathbf1\{\hat u_0(x)\ne u_0(x)\}$.
Exact recovery means $E=0$; for the moving-corner
experiment, equality is also verified entrywise.

\subsection{Spatial Redundancy and Decoding Rule}
\label{subsec:replica-decoding}

Frobenius replication provides a natural source of
spatial redundancy. For a fixed prime modulus $p$,
Lemma~\ref{lem:frobenius} gives, at $t^\ast=p^m$,
\[
u_{t^\ast}
=\sum_{a\in M}S_{p^m a}u_0-d\,u_0.
\]
When the supports of the nonzero terms are
pairwise disjoint, the resulting configuration
contains separated, coefficient-weighted copies
of the original seed.

Let $K$ denote the number of such copies, and
let $c_1,\ldots,c_K\in\mathbb F_p^\times$ be
their coefficients. After extracting and aligning
the corresponding tiles $v^{(1)},\ldots,v^{(K)}$,
we normalise them by
\[
w^{(j)}(x)=c_j^{-1}v^{(j)}(x),
\qquad j=1,\ldots,K.
\]
In the absence of noise, all normalised tiles
recover the original seed exactly.

For noisy configurations, we use pixelwise
consensus reconstruction:
\[
\hat u_0(x)=
\operatorname*{mode}_{1\le j\le K}w^{(j)}(x),
\]
with ties resolved by a fixed deterministic rule.
This procedure exploits the redundancy created
by replication without requiring the complete
evolution operator to be invertible.

For comparison, suppose that the normalised
replicas are corrupted independently, with the
correct symbol observed at each position with
probability $1-q$. In the binary case, for an odd
number $K$ of replicas, the probability of an
incorrect majority decision is
\[
q_{\mathrm{maj}}(K)
=\sum_{j=(K+1)/2}^{K}
\binom Kj q^j(1-q)^{K-j}.
\]

This expression illustrates the potential benefit
of independent replica errors in the binary case.
Our numerical experiment instead considers five
ternary replicas under the transmission-noise model
specified below.

\subsection{Transmission noise and replica-based reconstruction}
\label{sec:methods-noise}

We investigate the robustness of Frobenius replication
to errors occurring during transmission of an encoded
configuration. The encoding evolution is assumed to be
noiseless. Noise is introduced only after the final
encoded state has been obtained, rather than at
successive iterations of the Laplacian dynamics.

Let $S$ be a binary seed and let
$E=L_M^{t^\ast}S\pmod p$ denote its encoded state
at the Frobenius time $t^\ast=p^m$. Transmission
is modelled by $Y=E+\eta\pmod p$, where the
perturbation acts independently on individual
entries of the encoded state.

Each entry is selected for modification with
probability $\varepsilon$. A selected entry is
changed by adding a uniformly chosen nonzero
residue modulo $p$. Thus every selected entry
changes its value, while entries not selected
remain unchanged.

\paragraph{Replica extraction and reconstruction.}
For the diagonal neighborhood and $p=3$, the
Frobenius identity yields four diagonally
displaced copies of the seed with coefficient
$1$, together with a central copy whose
coefficient is $-4\equiv2\pmod3$.

At sufficiently large Frobenius times, these
five copies occupy disjoint regions. Their
positions and coefficients can therefore be
determined directly from the lattice operator.

The corresponding windows are extracted from
the received state $Y$. Each window is then
normalized by multiplying its entries by the
multiplicative inverse of its coefficient in
$\mathbb F_3$. This produces five candidate
reconstructions $R_1,\ldots,R_5$.

The final reconstruction $\widehat S$ is obtained
by pixelwise plurality voting:
\[
\widehat S(x)=
\operatorname*{arg\,max}_{v\in\mathbb F_3}
\sum_{j=1}^{5}\mathbf{1}\{R_j(x)=v\}.
\]
The voting procedure operates on the numerical
residue values, not on their graphical
representations.

\paragraph{Numerical experiment.}
The experiment uses a binary cat-shaped seed
of size $84\times84$, the diagonal neighborhood,
modulus $p=3$, and $t^\ast=243$ iterations.
Since $243$ exceeds the seed dimensions, the
five Frobenius replicas are spatially separated.

After noiseless encoding, independent symbol
errors are introduced once into the encoded
configuration. We compare reconstruction from
a single normalized replica with reconstruction
by voting across all five replicas.

The normalized Hamming error is defined by
\[
H(\widehat S,S)=
\frac{1}{|S|}
\sum_x \mathbf{1}\{\widehat S(x)\ne S(x)\},
\]
where $|S|$ denotes the number of entries in
the seed window.

For each realization, we record the error of
one normalized replica, the error after voting,
and whether the original seed was recovered
exactly. The numerical experiments were
implemented in Python using NumPy.

\subsection{Exact corner recovery under mixed moduli}
\label{subsec:corner_recovery}

The Frobenius identity describes the separation of spatial
contributions generated by a lattice operator, but it does
not guarantee that the entire configuration decomposes into
disjoint, unchanged copies of the initial seed. Overlapping
contributions may interfere, particularly under changes
of modulus.

A more precise statement is possible for an extreme corner
of the expanding configuration. We consider the diagonal
von Neumann neighborhood and show that its corner dynamics
has an exact, size-dependent period. This yields a recovery
guarantee for binary--ternary--binary schedules.

\paragraph{The moving corner operator.}
Let $S\in\{0,1\}^{D\times D}$ be an arbitrary binary seed
contained in a square of side length $D$. We consider the
diagonal Laplacian
\[
L_{\mathrm{diag}}=
S_{(1,1)}+S_{(1,-1)}
+S_{(-1,1)}+S_{(-1,-1)}-4I.
\]
At each iteration, the update is performed modulo the
prescribed integer, with residues represented by their
standard nonnegative values.

After $t$ iterations, let $C_t$ denote the $D\times D$
square at the upper-right corner of the expanding
propagation box. In particular, $C_0=S$. This window
moves one lattice unit upward and one unit to the right
at each iteration.

Let $U$ and $V$ denote the inward shifts on the corner
window, downward and leftward, respectively. Both are
nilpotent: $U^D=V^D=0$, and they commute.

A direct examination of the diagonal update shows that
the moving corner evolves autonomously. For a step
modulo a prime $p$, its operator is
\[
T_p=(I+U^2)(I+V^2)-4UV
\quad\text{over }\mathbb F_p.
\]
Thus no values from outside the moving corner window
are needed to determine its next state.

\begin{theorem}[Exact periods of the corner operator]
	\label{thm:exact_corner_periods}
	For the diagonal neighborhood on a $D\times D$ moving
	corner window, the exact orders of the operators
	$T_2$ and $T_3$ are
	\[
	P_2(D)=2^{\max\{0,\lceil\log_2(D/2)\rceil\}},
	\qquad
	P_3(D)=3^{\lceil\log_3D\rceil}.
	\]
	In particular, these are the smallest positive integers
	$n$ for which $T_2^n=I$ and $T_3^n=I$, respectively,
	on the entire corresponding state spaces.
\end{theorem}

\begin{proof}
	
	Since $T_p=I+N_p$, with $N_p$ nilpotent,
	the order of $T_p$ is a power of $p$.
	
	Over $\mathbb F_2$, the central term vanishes and
	$T_2=(I+U^2)(I+V^2)$. For $q=2^r$, the Frobenius
	identity gives
	\[
	T_2^q=(I+U^{2q})(I+V^{2q}).
	\]
	This equals $I$ whenever $2q\ge D$. If $2q<D$,
	the nonzero shift $U^{2q}$ cannot cancel against
	the distinct terms involving $V$. Hence
	$T_2^q\ne I$. Since the order divides the first
	such power of two, the stated formula for $P_2(D)$
	is exact.
	
	Over $\mathbb F_3$, let $q=3^s$. Frobenius yields
	\[
	T_3^q=
	I+U^{2q}+V^{2q}+U^{2q}V^{2q}-4U^qV^q.
	\]
	If $q\ge D$, all nonconstant terms vanish.
	If $q<D$, however, $U^qV^q$ is nonzero and cannot
	cancel against the other, distinct shift monomials.
	Thus $T_3^q\ne I$.
	
	Consequently, the first power of three yielding
	the identity is exactly $3^{\lceil\log_3D\rceil}$.
	The case $D=1$ is immediate for both operators.
\end{proof}

\begin{theorem}[Exact corner recovery under mixed moduli]
	\label{thm:mixed_corner_recovery}
	Let $S\in\{0,1\}^{D\times D}$ be an arbitrary binary
	seed, and let $P_2(D)$ and $P_3(D)$ be the periods
	given in Theorem~\ref{thm:exact_corner_periods}.
	
	Consider $a$ iterations modulo $2$, followed by
	$b$ iterations modulo $3$, and then $c$ further
	iterations modulo $2$, all using the diagonal
	neighborhood.
	
	If $P_3(D)\mid b$ and $P_2(D)\mid(a+c)$, then
	\[
	C_{a+b+c}=S.
	\]
	The equality is exact and entrywise: the upper-right
	corner contains the original binary seed, with
	unchanged pixel values and arrangement.
\end{theorem}

\begin{proof}
	After the first binary block, the corner state is
	$C_a=T_2^aS$, which has entries in $\{0,1\}$.
	
	Since $P_3(D)\mid b$, the ternary corner operator
	satisfies $T_3^b=I$. The ternary block therefore
	returns precisely the binary state $C_a$.
	Its reduction modulo $2$ leaves every entry unchanged.
	
	The final binary block consequently gives
	\[
	C_{a+b+c}=T_2^cT_2^aS
	=T_2^{a+c}S=S,
	\]
	where the last equality follows from
	$P_2(D)\mid(a+c)$.
\end{proof}

\paragraph{Recovery times and size-dependent scaling.}
The same argument applies at intermediate times
of the final binary block. Whenever $1\le j\le c$
and $P_2(D)\mid(a+j)$, we obtain
$C_{a+b+j}=S$, provided $P_3(D)\mid b$.

Thus exact corner recovery is guaranteed along an
arithmetic progression whose spacing is $P_2(D)$.
Both $P_2(D)$ and $P_3(D)$ are nondecreasing step
functions of the seed size, with jumps determined
by powers of two and three.

These periods are minimal for the operators acting
on all $D\times D$ configurations. They need not be
minimal for an individual seed, which may possess
additional structure or admit earlier returns.
Likewise, the divisibility conditions in
Theorem~\ref{thm:mixed_corner_recovery} are sufficient,
but are not asserted to be necessary for every
mixed-modulus recovery.

\paragraph{Scope and neighborhood geometry.}
The result concerns one extreme copy of the seed,
not the reconstruction of the complete evolved
configuration. Other regions may contain overlapping
contributions or additional modular patterns. Exact
corner recovery does not imply global invertibility
or cryptographic security.

By symmetry, corresponding statements hold for
the other three extreme corners of the diagonal
neighborhood.

The location of an extreme replica depends on the
neighborhood geometry. In particular, the standard
von Neumann neighbourhood propagates in four axial
extreme directions rather than toward the four
diagonal corners. More generally, extreme propagation
directions are related to the vertices of the convex
hull of the neighborhood mask.

Extending exact recovery results to other neighborhoods
requires an analysis of their corresponding moving
boundary regions.

\section{Results}\label{sec:results}

We first examine the revival structure of prime-modulus Laplacian dynamics, then 
quantify their entropy signatures, extend these observations to multi-prime 
compositions, and finally assess robustness under additive noise.

\subsection{Illustrative Chaos--Order Transitions}

\begin{figure}[h!]
	\centering
	\includegraphics[width=0.98\textwidth]{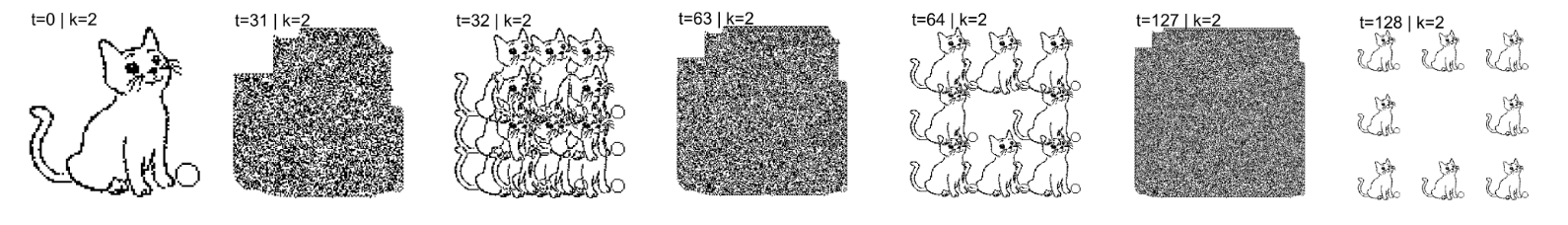}
	\caption{Binary Moore--Laplacian evolution of a
		``cat'' seed. Configurations at $t=31,63,127$
		precede the Frobenius times $t=32,64,128$,
		respectively. The latter exhibit algebraically
		determined replication, with spatially separated
		copies visible at $t=128$.}
	\label{fig:cat-chaos-order}
\end{figure}

We begin with a visual illustration of Frobenius-driven
replication. Figure~\ref{fig:cat-chaos-order} shows the
evolution of a binary ``cat'' seed under the Moore
Laplacian. The configurations become increasingly
dispersed and visually complex, but undergo abrupt
reorganisation at distinguished iterations.

The contrast is particularly clear between the states
immediately preceding the binary Frobenius times and
those obtained at $t=32,64,128$. At these times,
the algebraic identity of Lemma~\ref{lem:frobenius}
reduces the evolution to a superposition of translated
copies of the initial seed.

As the translation distances increase, overlapping
copies eventually become spatially separated.
The configuration at $t=128$ illustrates this
separated-replication regime.

\subsection{Constant-Modulus Baseline}
\label{sec:baseline}

Replication under a fixed modulus has been examined
in our earlier work
\cite{MNK2025laplacian,MNK2025fractal}.
Here we recall the properties needed for the
comparison with mixed-modulus evolution.

For a prime $p$, the Frobenius identity guarantees
a replication formula at times $t=p^m$.
The number and coefficients of the surviving
copies depend on the neighbourhood mask and
the reduction modulo $p$, whereas their spatial
separation depends on the size of the initial seed.

For the Moore neighbourhood, the Laplacian has
eight neighbouring terms and a central coefficient
$-8$. Hence, at $t=p^m$, the evolution consists
of eight translated contributions and, unless
$p$ divides $8$, an additional central contribution.

If the seed is contained in a rectangle of width
$w$ and height $h$, a sufficient condition for
these contributions to have disjoint supports is
$p^m\geq\max\{w,h\}$.
Consequently, the first guaranteed separated
Frobenius time for this mask is
\[
T_{\mathrm{sep}}
=
\min\{p^m:p^m\geq\max\{w,h\}\}.
\]
This bound concerns the geometry of the translated
supports; particular seeds may separate earlier.

Table~\ref{tab:small-big-periods} summarises the
prime-power scales and the corresponding number
of nonzero contributions for the Moore mask.

\begin{table}[h!]
	\centering
	\caption{Frobenius replication for the Moore
		Laplacian at prime modulus $p$.
		The number of contributions assumes spatial
		separation of the corresponding supports.}
	\label{tab:small-big-periods}
	\begin{tabular}{c c c c}
		\toprule
		$p$ & Frobenius times & Central coefficient
		& Nonzero copies \\
		\midrule
		$2$ & $2^m$ & $0$ & $8$ \\
		$3$ & $3^m$ & $1$ & $9$ \\
		$5$ & $5^m$ & $2$ & $9$ \\
		\bottomrule
	\end{tabular}
\end{table}

\begin{figure}[h!]
	\centering
	(a)\includegraphics[width=0.80\textwidth]{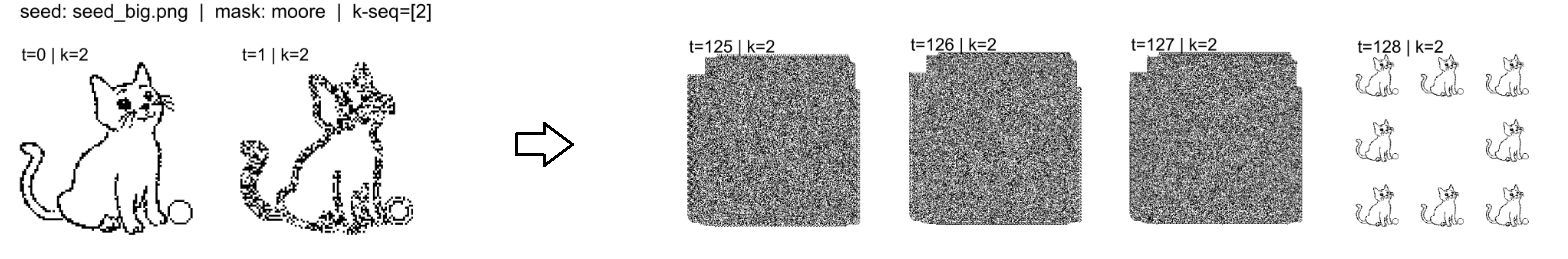}\\[2mm]
	(b)\includegraphics[width=0.80\textwidth]{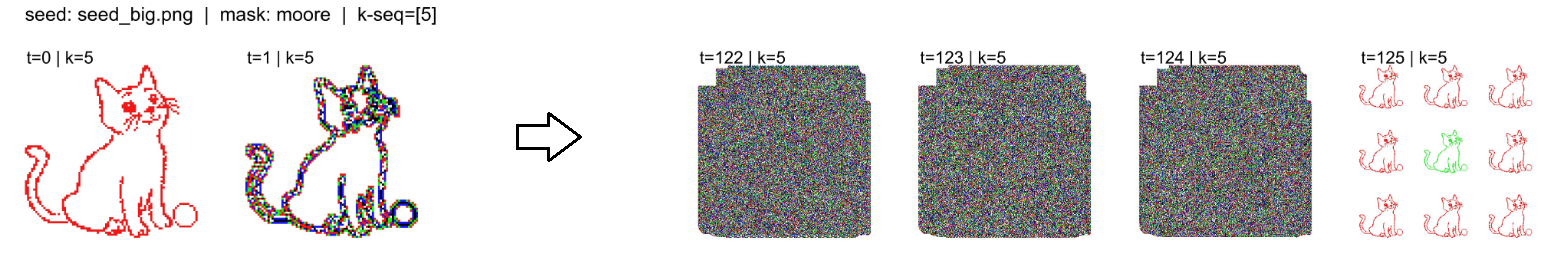}

\caption{Frobenius-driven replication of a
	natural-image silhouette under the Moore Laplacian:
	(a) binary evolution, with eight separated copies
	at $t=128$; (b) quinary evolution, with nine
	separated copies at $t=125$.
	The central copy in (b) carries coefficient
	$-8\equiv2\pmod5$, whereas the surrounding
	copies retain coefficient $1$.
	The preceding configurations illustrate the
	dispersive evolution immediately before replication.}

	\label{coding}
\end{figure}

Figure~\ref{coding} illustrates the phenomenon
for binary and quinary evolution of the cat seed.
The relevant Frobenius scales include $t=128$
for $p=2$ and $t=125$ for $p=5$.
The resulting patterns demonstrate that the same
algebraic mechanism can produce different
replication geometries and symbol values.

These constant-modulus examples provide a
reference for the more general, order-dependent
modular sequences considered below.

\subsection{Entropy Dynamics}
\label{sec:entropy}

\begin{figure}[h!]
	\centering
	\includegraphics[width=0.90\textwidth]{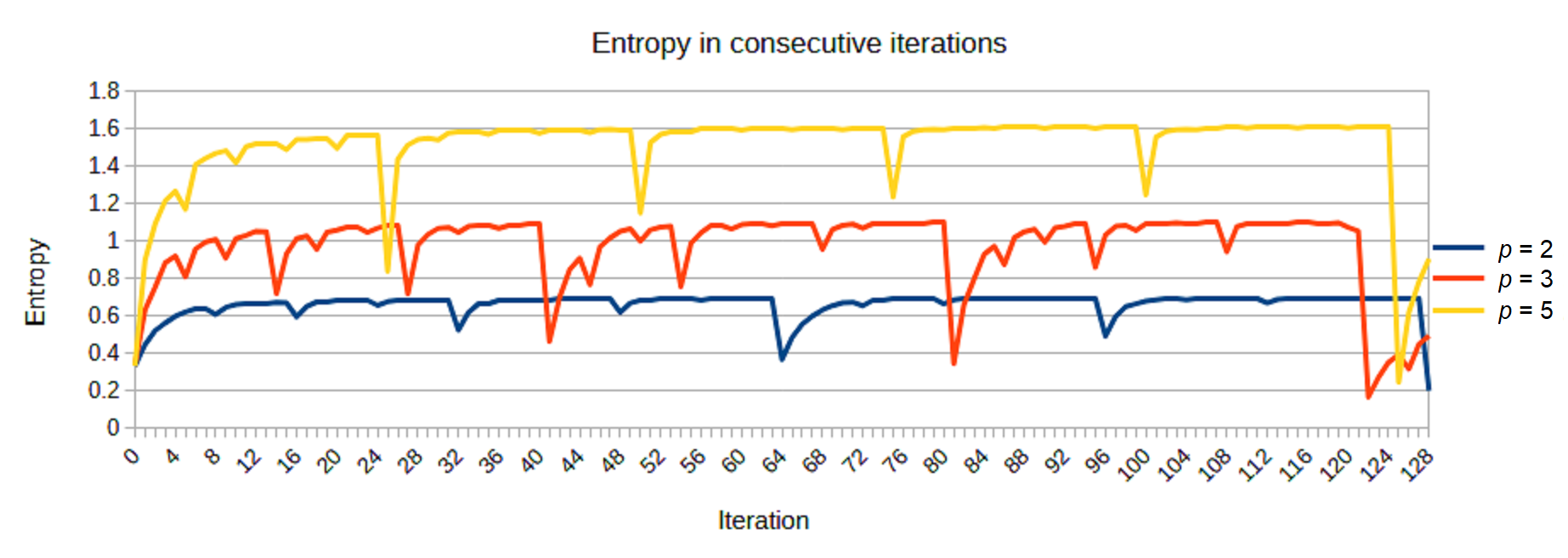}
	\caption{Shannon entropy $H_t$ for the
		Moore--Laplacian evolution of the cat seed
		under prime moduli $p=2,3,5$, over
		$128$ iterations. The corresponding Frobenius
		times include $t=128$, $81$, and $125$,
		respectively. Entropy fluctuations provide
		a statistical description of the evolving
		configurations but are not, by themselves,
		a criterion for exact replication.}
	\label{fig:entropy-2-3-5}
\end{figure}

To complement the geometric observations,
we examine the Shannon entropy $H_t$ defined
in Section~3.3. Figure~\ref{fig:entropy-2-3-5}
shows its evolution for the cat seed under
constant prime moduli $p=2,3,5$.

In all three cases, the entropy increases during
the initial dispersive evolution and subsequently
exhibits pronounced fluctuations. Several local
minima occur near configurations displaying
strong spatial organisation.

The entropy scale depends on the alphabet size,
with the theoretical bound $H_t\leq\log p$.
Thus entropy values for different moduli should
be interpreted relative to their respective
symbol alphabets.

The Frobenius times $t=128$ for $p=2$,
$t=81$ for $p=3$, and $t=125$ for $p=5$
provide algebraically distinguished reference
points for interpreting the entropy trajectories.

However, a replication event does not necessarily
correspond to a unique or global entropy minimum.
The measured entropy also depends on the symbol
frequencies and on the spatial window used in its
calculation. Conversely, low entropy alone does
not establish that a configuration consists of
translated copies of the initial seed.

We therefore use entropy as a complementary
indicator of spatial organisation, while
identifying exact Frobenius replication through
the algebraic identities established in
Section~2.2.

\subsection{Robustness to transmission errors}
\label{sec:noise}

We evaluated replica-based reconstruction after
independent symbol errors were introduced into
the encoded configuration. The experiment used
a binary cat-shaped seed of size $84\times84$,
the diagonal neighborhood, modulus $3$, and
the Frobenius time $t^\ast=243$.

At this time, the encoded state contains five
spatially separated copies of the seed, whose
coefficients can be normalized before voting.
We tested a transmission error probability of
$\varepsilon=0.01$, corresponding to a 1\%
probability of modifying each transmitted symbol.

The results are presented in
Table~\ref{tab:transmission_noise}.
The normalized Hamming error of a single
replica ranged from $0.010629$ to $0.011905$,
with a mean of $0.011243$.

After pixelwise voting across the five normalized
replicas, the reconstructed seed agreed exactly
with the original in all three trials.

\begin{table}[htbp]
	\centering
	\caption{Reconstruction from five separated
		Frobenius replicas under independent transmission
		noise ($\varepsilon=0.01$).
		Normalized Hamming errors are reported for
		one replica and after pixelwise voting.}
	\label{tab:transmission_noise}
	\begin{tabular}{ccc}
		\toprule
		Trial & Single replica & Replica voting \\
		\midrule
		1 & 0.010629 & 0.000000 \\
		2 & 0.011905 & 0.000000 \\
		3 & 0.011196 & 0.000000 \\
		\midrule
		Mean & 0.011243 & 0.000000 \\
		\bottomrule
	\end{tabular}
\end{table}

These results demonstrate that spatial redundancy
generated by Frobenius replication can be used
to correct transmission errors. In the tested
realizations, voting eliminated all symbol errors
present in the reconstructed seed.

The mechanism relies on the availability of
several separated representations of the same
initial configuration. Since transmission noise
is introduced independently at different
spatial locations, the corresponding replica
errors can be reduced by voting.

The experiment does not imply that every
transmission error pattern can be corrected.
In particular, sufficiently extensive or
correlated errors may affect several replicas
at the same pixel position. The observed exact
recovery should therefore be interpreted as
a numerical demonstration under the specified
independent symbol-error model, rather than
as a general guarantee of error-free decoding.


\subsection{Exact corner recovery under mixed moduli}
\label{subsec:corner_results}

\begin{figure}[htbp]
	\centering
	\includegraphics[width=\textwidth]{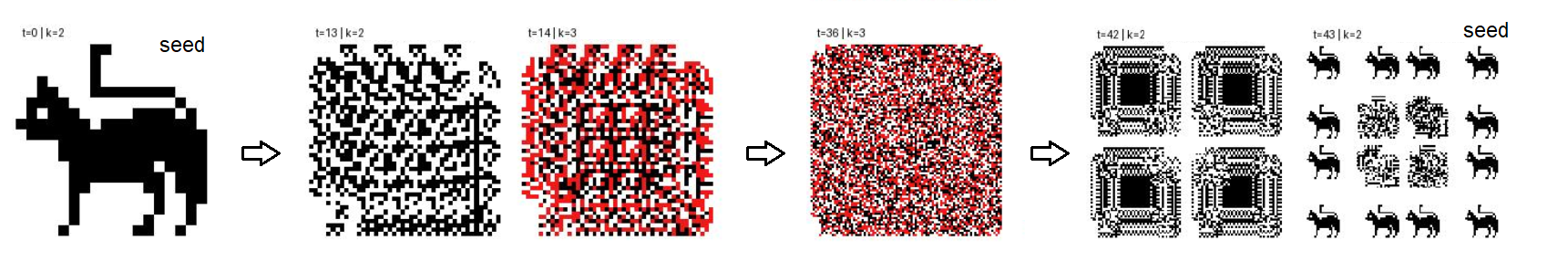}

\caption{Exact corner recovery under a
	binary--ternary--binary schedule. A cat-shaped
	binary seed evolves under the diagonal von Neumann
	neighborhood through 13 iterations modulo 2,
	27 iterations modulo 3, and three final iterations
	modulo 2. Selected states are shown at
	$t=0,13,14,36,42$, and $43$.
	At the final time, recognizable copies of the seed
	appear along the outer boundary of the configuration.
	In particular, the upper-right corner contains an
	exact copy of the original seed, while the interior
	contains additional, partly overlapping structures.}

	\label{fig:corner_recovery}
\end{figure}

The exact corner recovery mechanism established in
Theorem~\ref{thm:mixed_corner_recovery} is illustrated
in Fig.~\ref{fig:corner_recovery}. We consider a binary
cat-shaped seed evolving under the diagonal von Neumann
neighborhood, with the schedule
$2^{[13]}3^{[27]}2^{[3]}$, where the bracketed
exponents denote consecutive iterations under the
corresponding modulus.

The first binary block produces an extended, irregular
configuration ($t=13$). The subsequent ternary block
introduces an additional nonzero residue and generates
a more intricate spatial pattern, illustrated at
$t=14$ and $t=36$. After returning to modulo 2,
the configuration at $t=42$ exhibits a pronounced
regular structure, followed by the appearance of
recognizable seed copies at $t=43$.

The recovery conditions are satisfied: the ternary
block has length $b=27$, while the binary blocks
have combined length $a+c=13+3=16$. At $t=43$,
the right upper corner windows reproduce the
original binary seed exactly. This agreement was
verified entrywise on the modular state arrays,
rather than inferred from the rendered images.

The interior of the final configuration, however,
contains additional and partly overlapping patterns,
including distorted seed-like structures. Thus the
experiment illustrates the distinction between
exact recovery at the propagation extremes and
global replication of the initial configuration.

\section{Applications}
\label{sec:applications}

\subsection{Pattern generation}
At prime-power times, Frobenius identities express the
evolved state as a superposition of translated,
coefficient-weighted seed copies. When their supports
are disjoint, the result provides a deterministic way
to generate repeated spatial motifs. The geometry of
the mask and the initial support determine when these
motifs can be resolved as separate copies.

\subsection{Information representation and recovery}
For a fixed prime modulus, an intermediate configuration
$u_s=L_M^s u_0$ can be evolved forward to a known
Frobenius time $t^*=p^m$. If that state contains a
separated replica with nonzero coefficient, the seed
can be recovered by extracting and normalising the tile.
This is a forward recovery procedure and does not
assert invertibility of $L_M$.

Multiple separated replicas also permit consensus
reconstruction in the presence of errors, but correlated
noise can limit its effectiveness. For the diagonal
neighborhood, Theorem~\ref{thm:mixed_corner_recovery}
provides a different, mixed-modulus guarantee: the
original seed reappears exactly in a moving extreme
corner under stated divisibility conditions. Neither
mechanism by itself establishes encryption security.

\section{Discussion}

The results distinguish three related properties:
algebraic replication at prime-power times, geometric
separation of the contributing copies, and exact recovery
of the initial seed. The Frobenius identity establishes
the first property, while the second depends on the mask
and seed support. Recovery from a separated replica is
exact after normalisation by its nonzero coefficient,
but it is not equivalent to reversing the full evolution.

The moving-corner analysis yields a complementary result
for mixed-modulus dynamics. For the diagonal neighborhood,
we obtained exact, size-dependent periods of the binary
and ternary corner operators and sufficient conditions
for recovering an arbitrary binary seed after a
binary--ternary--binary schedule. This statement concerns
extreme windows rather than the entire final pattern.
The observed stepwise increase of the guaranteed periods
with seed size follows from nilpotency and Frobenius,
not from the appearance of a particular silhouette.

Experiments with local perturbations illustrate finite
propagation, whereas the accumulated-noise experiments
probe a different question: the effectiveness of
consensus reconstruction when errors can become correlated.
The reported noise thresholds are empirical and depend on
the chosen seed, grid, mask and injection protocol; no
universal error-correction bound follows.

The visual complexity of intermediate configurations
should not be interpreted as randomness or security.
Likewise, the exact corner result does not characterise
all successful mixed-modulus schedules. Determining
minimal recovery times for specific seeds, extending
the construction to other extreme directions, and
testing combinations of spatial and temporal redundancy
remain open directions. Multiple temporal snapshots
could be decoded separately and combined, but their
potential advantage has not been evaluated here.

\section{Conclusion}

We studied Frobenius revivals in modular Laplacian
cellular automata, separating algebraic replication,
spatial separation and information recovery. At
prime-power times, the evolution has an exact
coefficient-weighted replication formula; disjoint
copies allow noiseless recovery by tile extraction
and normalisation.

For the diagonal neighborhood we also determined the
exact periods of moving corner operators on square
windows and proved sufficient conditions for the exact
recovery of any binary seed under a
binary--ternary--binary schedule. This result explains
a controlled form of revival despite changes of modulus,
without implying reconstruction of the full state.

The redundancy of separated Frobenius replicas
also enables correction of transmission errors.
In the reported numerical experiment, voting
over five replicas recovered the original seed
exactly in all three trials under 1\%
independent symbol noise.

Numerical illustrations show both the spatial patterns
associated with these identities and the possibilities
and limitations of reconstruction under perturbations.
The results provide an algebraic basis for studying
structured pattern generation and further questions
about mixed-modulus recovery.

\section*{Statements}

\paragraph{Funding.}
This research received no specific grant from any funding agency
in the public, commercial, or not-for-profit sectors.

\paragraph{Conflict of interest.}
The author declares no conflict of interest.

\paragraph{Ethical approval.}
Not applicable. This study did not involve human participants
or animals.

\paragraph{Data and code availability.}
The Python scripts, input seeds, and supporting numerical
data are publicly available in the Frobenius Dataset
deposited on Figshare\\
\url{https://figshare.com/articles/dataset/_/34338891}

\paragraph{Author contributions.}
Małgorzata Nowak-Kępczyk: Conceptualization,
Methodology, Formal analysis, Software,
Investigation, Visualization, Writing --
original draft, Writing -- review and editing.

\bibliographystyle{apalike}

\bibliography{references}   
\end{document}